\documentclass[epj]{svjour}
\usepackage{graphicx}
\begin{document}
\title{Event shape sorting
\thanks{Supported in parts by APVV-0050-11, VEGA 1/0469/15 (Slovakia) and 
M\v{S}MT grant  LG13031, SGS15/093/OHK4/1T/14 (Czech Republic).}%
}
\author{Renata Kope\v{c}n\'a\inst{1} \and Boris Tom\'a\v{s}ik\inst{1,2}
}                     
%
%
\institute{FNSPE, Czech~Technical~University in Prague, B\v{r}ehov\'a~7, 11519~Prague,
Czech~Republic \and 
Univerzita~Mateja~Bela, Tajovsk\'eho~40, 97401~Bansk\'a~Bystrica, Slovakia}
\date{Received: \today / Revised version: date}
%
\abstract{
We propose a novel method for sorting events of multiparticle production according to 
the azimuthal anisotropy of their momentum distribution. Although the method is quite general, 
we advocate its use in analysis of ultra-relativistic heavy-ion collisions where large number 
of hadrons is produced. The advantage of our method is that it can automatically sort 
out samples of events with histograms that indicate similar distributions of hadrons.
It takes into account the whole measured histograms with all orders of anisotropy 
instead of a specific observable (e.g.\ $v_2$, $v_3$, $q_2$). 
It can be used for more exclusive experimental 
studies of flow anisotropies which are then more easily compared to theoretical calculations. 
It may also be useful in the construction of mixed-events background for 
correlation studies as it allows to select events with similar momentum distribution. 
} 
\maketitle

\section{Introduction}

Hot matter which is created in ultrarelativistic heavy-ion collisions expands very fast 
in both longitudinal and transverse directions \cite{WPphenix,WPphobos,WPbrahms,WPstar}. 
The expansion is always anisotropic. 
This is true even in most central collisions, where one would
expect symmetry in azimuthal angle due to  circular shape of the initial overlap 
region \cite{CMSvn}. In non-central collisions, 
the overlap of the two colliding nuclei has elliptic shape and thus naturally second-order
(and higher even orders) anisotropy in the fireball expansion builds up. 
On top of this---at any centrality---energy deposition in the interactions of the incoming 
partons fluctuates and this leads to all orders of anisotropy in the transverse 
expansion of the fireball. Therefore, even within carefully selected centrality class
flow anisotropies vary from event to event \cite{CMSvn,Milov}. 

Even more degrees of freedom are  offered by collisions of deformed nuclei, 
like uranium. There, the 
initial anisotropy will also depend on the way how the colliding nuclei are oriented. 

This makes the comparison of experimental data to theoretical simulations more complicated,
because one has to take into account that every event starts with different initial conditions
and evolves differently. 
Simulations are compared to data in order to pin down 
the properties of the matter which is being modelled.
Initial conditions are unknown, however, although recent hydrodynamic results 
indicate that their fluctuations can be directly 
mapped onto measured fluctuations of hadron 
distributions \cite{niemi,gale}.

Comparison of theory to data must be done with great care
so that the spectra of theoretical and experimental fluctuations match each other.
Experimentally, events are distributed into centrality classes according to multiplicity.
There is a problem with this procedure on the side of  theory if very narrow centrality
class is demanded, e.g.\ ultra-central collisions corresponding to 0--0.2\% centrality.
Events with the same multiplicity may evolve from initial states with different 
impact parameters\footnote{%
This can be seen, e.g., in Fig.~2 of  \cite{ALICE_centr}, where the procedure is explained
that is used by ALICE collaboration to determine centrality. 
Within the used Monte Carlo 
Glauber model if the impact parameter is fixed, then the number of participants 
may still fluctuate. 
On the other hand, from the overlap of centrality classes in $N_{\mathrm{part}}$ histogram 
it is clear that fixed  $N_{\mathrm{part}}$ corresponds to an \emph{interval} of impact 
parameters. As the experimental multiplicity is determined from multiplicity in a chosen detector, 
there is yet another source of fluctuation that comes from the uncertainty between 
multiplicity and $N_{\mathrm{part}}$. In summary, if we fix the multiplicity---even in perfectly 
spherical nuclei like Pb or Au---there is still some freedom for the impact parameter to fluctuate 
which we estimate of the order 1~fm. Even more fluctuations can be expected in collisions
of non-spherical nuclei, like U. 
}.
Moreover, all those events would  
differ by the quantum fluctuations in initial energy and momentum deposition
\cite{Shen:2015qta}. 
Therefore, events from a class selected according to multiplicity-based centrality 
may have evolved from different initial conditions and experienced quite different
evolution history. It would be useful if there was a more selective method to choose 
events that are more likely to have evolved similarly. 
In addition to this, in collisions of deformed nuclei the multiplicity itself is 
definitely not sufficient
selection criterion since the same multiplicity may result from events with very different
initial \emph{orientations} of the 
colliding nuclei and thus very different flow patterns. 
Again, the situation calls for a selection method like the one presented here. 

It would thus be advantageous if one could select a collection of events  among  
all measured events (which may or may not belong to the same centrality class) which 
show very similar distribution of the produced hadrons. For such events one can 
assume that they also evolved similarly.

The word ``similar'' when talking about distributions or histograms can be understood in
layman's terms so that they  have similar shapes when you 
rotate them appropriately. There is
a possibility how to quantify this with the help of the distance measure 
in the Kolmogorov-Smirnov
test, but this will not be used here
because it does not automatically provide a way for sorting. 
We rather use the Bayesian framework where
we basically ask the question: how would the events be grouped, based on the shapes 
of their histograms.

A method aiming for such event selection 
has been proposed in \cite{ese} and is commonly referred to as
``Event Shape Engineering''. It usually employs the size of the flow vector $q_n$  
defined on a selection 
of hadrons from a given event (usually referred to as subevent) as
\begin{equation}
\vec q_n = \frac{1}{\sqrt{M_s}} \left ( \sum_{i=1}^{M_s} \cos(n\phi_i) , 
\,\sum_{i=1}^{M_s} \sin(n\phi_i)
\right ) \,  ,
\end{equation}
where $M_s$ is the multiplicity of the subevent and $\phi_i$ are the azimuthal angles of the 
individual hadrons  from  the subevent. 
Events are then selected based on their values of $q_n$ (in most cases $q_2$).
However, only the subevents \emph{not}
used in $q_n$ determination can be used in further studies in order to avoid bias. 

It is not clear, moreover, whether selecting events according to the value of $q_n$
provides the best possible selection method aimed at collecting similar events.
It actually may not be the case, as we demonstrate below.

In this paper we propose the use of a novel method for comparing, sorting and 
selecting events according to \emph{similarity} with each other. The method is adopted 
from \cite{Lehmann,Lehmann:2007pv} and was previously used in a different context \cite{Jackson}. 
Its uniqueness consists in not fixing a single observable which would then be used 
for sorting of the events. It rather compares complete histograms (e.g.\ 
azimuthal angle distributions)
of individual events and it sorts the events in such a way that events with similar histogram
shapes
end up close together.  After such a sorting has been performed one simply selects 
similar events just by choosing a group of events which follow each other in the created series.

On the selected groups of events one could measure various observables ($v_n$'s, 
$q_n$'s, radial flow, temperature, \dots) which should fluctuate much less than in the 
whole measured event sample. 

A natural application of the method is in construction of  correlation functions. 
There, one often needs a reference distribution which is constructed via the mixed 
events technique. If events used in mixing are different, this may introduce unwanted 
artificial effects into the correlation function. Therefore, the mixed events sample is 
always constructed with aligned event planes. (For application at 
intermediate energy nuclear collisions see e.g.\ \cite{Kampfer:1993zz,Kotte:1995zz}). 

With the help of the proposed method it would be interesting to perform femtoscopic 
studies where oscillations of radii in azimuthal angle 
in both second and third order together at the same time should be visible. 

We comment more on the applications in the Outlook section.

In the next Section we shall explain the method and in Section \ref{s:illus} we illustrate 
its use on Monte Carlo data from a toy model. 
The method is applied on more realistic Monte Carlo data generated by a
transport model in Section \ref{s:ampt}. We conclude in Section~\ref{s:conc}
and give an Outlook about possible applications of the method and its next development.


\section{The method}


Suppose that we have a sample consisting of a 
large number of events. Initially, we can sort and divide those 
events into $N$ percentiles according to the value of a single observable  
which can be measured 
in every event. This can be the value of $q_2$, $v_2$, multiplicity or any other 
observable. Generally, we shall refer to this observable as $Q$. 
For the sake of clarity let us explain the method 
with a particular choice: choose $N=10$ event bins (deciles)
and the observable according to which we sort data is 
$Q =q_2 =|\vec q_2|$. Let us stress at this 
point that the method is universal and will not depend on the choice of $N$ and $Q$.

In each of the event bins we can now produce the histogram of hadron distribution 
in azimuthal angle summed over all events in the event bin. There is no physics 
in how the two nuclei are oriented when they collide, and so
we have a free choice of how to rotate individual events before adding them 
to the angle histogram. For the introduction of the method let us align each event
according to the second order event plane. Note, however, that the choice of alignment 
is a sensitive issue which we shall discuss later.

Thus each event is characterized by the bin record of its distribution in azimuthal 
angle\footnote{%
We denote $n_i$ the number of particles in angle bin $i$ and the whole record of an event 
is referred to with the help of braces. Thus summation of all angle bin entries gives the event multiplicity$\sum_i n_i = M$.}
$\{ n_i \}$ and belongs to one of the $N$ event bins which we number with $\mu$. 
(We shall use Latin letters for angle bins and
Greek letters for event bins.)


\subsection{Basic relations}

Imagine that we take a random event from our big sample and ask an unbiased
observer, in which event bin he or she thinks that this event belongs. More specifically, we can 
ask the question in the framework of Bayesian probability: What is the probability\footnote{%
Obviously, the event must belong to one of the event bins so that the 
probabilities must be 
normalised 
\[
\sum_{\mu = 1}^N P(\mu|\{ n_i \} ) =1\,  .
\]
}%
$P(\mu|\{n_i\})$ that an 
event with bin record $\{ n_i \}$ belongs to the event bin $\mu$?

This probability will be evaluated with the help of Bayes' theorem 
\begin{equation}
\label{e:bayes}
P(A|B) = \frac{P(B|A) P(A)}{P(B)}\,  ,
\end{equation}
where $P(A|B)$ is the conditional probability of the event $A$ given event $B$. The probability 
of event $B$ can be determined
\begin{equation}
\label{e:bayes_den}
P(B) = \sum_A P(B|A) P(A)\,  ,
\end{equation}
where the sum runs over all possible events $A$. Definitions of symbols $P(B|A)$ and 
$P(A)$ are analogical. 

With the help of Bayes' theorem we can express
\begin{equation}
\label{e:bay_appl}
P(\mu|\{n_i\}) = \frac{P(\{n_i\}|\mu) P(\mu)}{P(\{n_i\})}\,  .
\end{equation}
Here, $P(\{n_i\}|\mu)$ is the probability that if one randomly draws an event from the
distribution function given by average histogram of event bin $\mu$, the result will be
the bin record $\{n_i\}$. The prior $P(\mu)=1/N$ takes the value of 0.1 now. 
The denominator contains the overall probability of drawing the event $\{n_i\}$ from any of
the event bins. It can be determined according to eq.~(\ref{e:bayes_den})
\begin{equation}
\label{e:suma}
P(\{ n_i \})  = \sum_{\nu=1}^N P(\{n_i\}|\nu) P(\nu)\,  .
\end{equation}
The advantage of using the latter formula comes from the fact that we are able to 
determine $P(\{n_i\}|\nu)$ for 
each bin record $\{n_i\}$ and every event bin $\nu$
\begin{equation}
\label{e:kombin}
P(\{n_i\}|\nu) = M! \prod_i \frac{P(i|\nu)^{n_i}}{n_i!}\,  . 
\end{equation}
Here $M$ is event multiplicity, 
the product goes over all angle bins $i$ and $P(i|\nu)$ is the conditional 
probability that random particle falls into angle bin $i$ given that the event to which 
it belongs stems from event bin $\nu$. It can be determined when we take
the number of particles from all events in event bin $\nu$ falling into  
angle bin $i$. This number is divided  by the total number of all particles 
from events in event bin $\nu$, denoted $M_\nu$
\begin{equation}
P(i|\nu) = \frac{n_{\nu,i}}{M_\nu}\,  .
\end{equation}

When the formula (\ref{e:kombin}) is inserted into (\ref{e:suma}) and 
(\ref{e:bay_appl}) we obtain the practically usable relation from which the large factorials 
drop out 
\begin{equation}
\label{e:e8}
P(\mu|\{n_i\}) = \frac{\prod_i P(i|\mu)^{n_i} P(\mu)}{\sum_\nu \prod_i P(i|\nu)^{n_i} P(\nu)}\, .
\end{equation}

With the help of this conditional probability we can determine for an event with 
angle bin record $\{ n_i \}$ its 
\emph{mean event bin number}
\begin{equation}
\label{e:meanmu}
\bar \mu = \sum_\mu \mu P(\mu|\{n_i\})\,  .
\end{equation}


\subsection{The algorithm}
 
Now we are able to describe the algorithm which is used for sorting of the events. 
\begin{enumerate}
\item 
First, all events are sorted according to the observable $Q$. 
\item
Events are divided into $N$ event bins according to current sorting.
\item 
For each event and all event bins the probability\linebreak 
$P(\mu|\{n_i\})$ is determined that 
the event with record $\{n_i\}$ belongs  to event bin 
$\mu$. The mean event number $\bar \mu$ is calculated for each event according to 
(\ref{e:meanmu}).
\item 
Events are sorted again according to their values of $\bar \mu$. 
\item
Events are divided into $N$ event bins according to current sorting. 
\item
If the new sorting changed assignment of any events into event bins,
the algorithm returns to step 3. Otherwise 
it converged.
\end{enumerate}
The construction of the algorithm is such that once converged, the events are sorted so
that those ending up close to each other are characterized by similar angular histograms. 
This is the best possible experimental approach to the selection of events that have 
undergone  similar evolution. This is as good
working definition of what  similar events are as it can be. 

In the present 
formulation the average histograms are more strongly determined by 
high-multiplicity events. 
Note, however, that the algorithm is independent of event multiplicity.
Of course, if in particular physics
analysis certain multiplicity is demanded, one can easily use it for event selection and then 
use the present method for more refined selection of the events. 

The final sorting of events also does not depend on the initial sorting. Hence, this method 
can also be used for a judgment if the particular observable $Q$, 
e.g.\ $q_2$ or $v_2$, is a good measure for selecting similar events. 
If the initial ordering according to $Q$ is correlated with  the final ordering, then 
the observable $Q$ is good for this purpose.
This may not always be the case, as we will show later. 

Although the final result does not depend on initial ordering, a good initial ordering 
can lead to faster convergence of the algorithm. 

There is a caveat, however, which has been discussed only shortly so far. 
The algorithm works well for sorting histograms 
in observables that are not periodic, e.g.\ rapidity. However, the azimuthal angle 
of particle momentum is a periodic observable, i.e.\ we can always arbitrarily rotate the 
events. Practically, two angular histograms 
may be almost identical when they are both aligned properly, but might appear 
quite different for the  proposed algorithm if they are rotated in random directions. 
Thus the way how events are aligned initially plays a crucial role and we observed that 
it strongly biases the final sorting of the events. At the moment we do not have 
recommendation for an automatic algorithm which would align the events in the best way. 
Instead, we tested a few  reasonable choices for the initial alignment in our toy model 
studies. 
With the simple toy model, 
quite naturally that initial rotation aligning the second-order 
event planes $\psi_2$ will  lead to sorting characterized by $v_2$. Analogically, with aligning 
the third-order event planes $\psi_3$, final sorting is given by $v_3$. In the next 
Section we present results which take into account a combination of both these 
orderings.
That feature was, however, weaker in AMPT-generated events.


\section{Illustration of the method}
\label{s:illus}

\subsection{Elliptic flow}

Let us first demonstrate the action of the sorting algorithm on a simple case 
of events with only first- and second-order anisotropic flow. We generated 
the azimuthal angles of  pions from the distribution
\begin{equation}
P_2(\phi) = \frac{1}{2\pi} \left ( 1 + 2v_1\cos(\phi - \psi_1) 
+ 2v_2 \cos(2(\phi - \psi_2)) \right )\, .
\end{equation}
The parameters $v_1$ and $v_2$ depend on the multiplicity of the event  $M$ as
\begin{equation}
v_n = a_n M^2 + b_n M + c_n\,  ,
\label{e:vnM}
\end{equation}
and  parameters $a_n$, $b_n$, and $c_n$ for each $n$ can be found 
in Table~\ref{t:params}. They have been determined from experimental data as 
we discuss in Section \ref{s:aniflow}. In addition to eq.~(\ref{e:vnM}), flow anisotropy
parameters $v_n$ are Gaussian-smeared with a width of 0.25. For each event, 
the directions of event planes $\psi_1$ and $\psi_2$ are random
and uncorrelated.

\begin{table}
\caption{Parameters used for generating multiplicity dependent $v_n$.
\label{t:params}}
\begin{center}
\begin{tabular}{cccc}
\hline
$n$ & $a_n\times 10^{8}$ & $b_n\times 10^{5}$ & $c_n$ \\
\hline\hline
1 & 0 &  0.01667 & --0.000680 \\
2 & --7.099 & 20.06 & 0.07874 \\
3 & --2.083 &  6.658 & 0.04236  \\
4 & --96.38 &  2.621 & 0.04897 \\
5 & --71.76 &  2.236 & 0.01673 \\
\hline
\end{tabular}
\end{center}
\end{table}

We generated 5000 events with multiplicities between 300 and 3000. 
Directed flow is practically negligible and the dominant anisotropy is second-order. 
Thus the most reasonable choice of initial event rotation is the alignment of 
second-order event planes defined from the generated Monte Carlo data for 
each event via
\begin{equation}
q_2 e^{2i\psi_2} = \sum_{j=1}^{M} e^{2i\phi_j}\, .
\end{equation}

To show the power of the method we first ordered the events fully randomly. 
The algorithm converged after 65 iterations. 
\begin{figure}
\centerline{\includegraphics[width=0.42\textwidth]{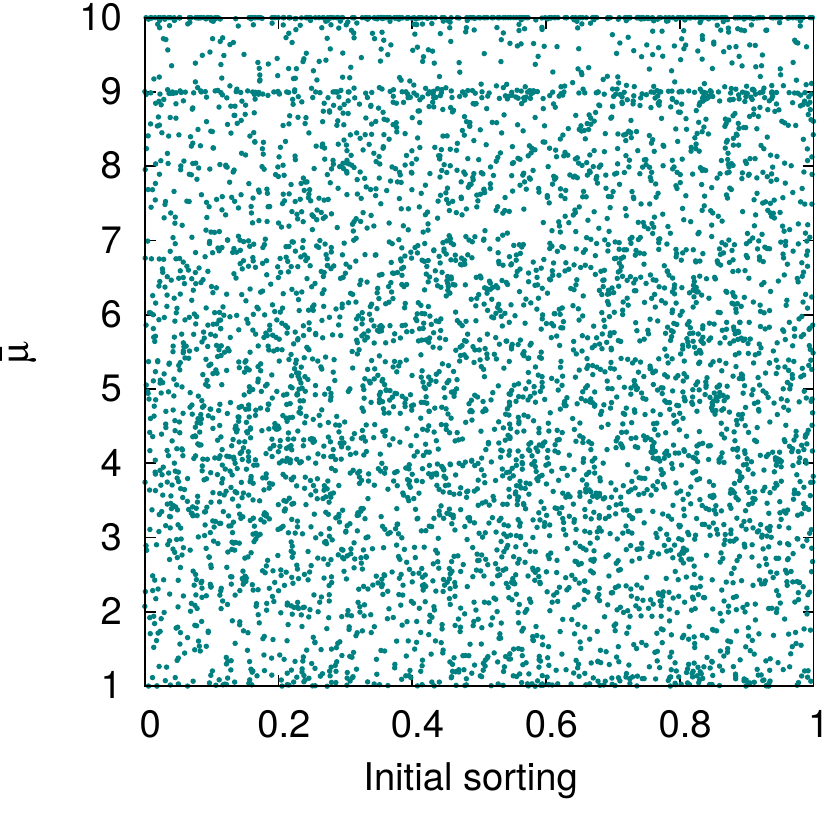}}
\centerline{\includegraphics[width=0.42\textwidth]{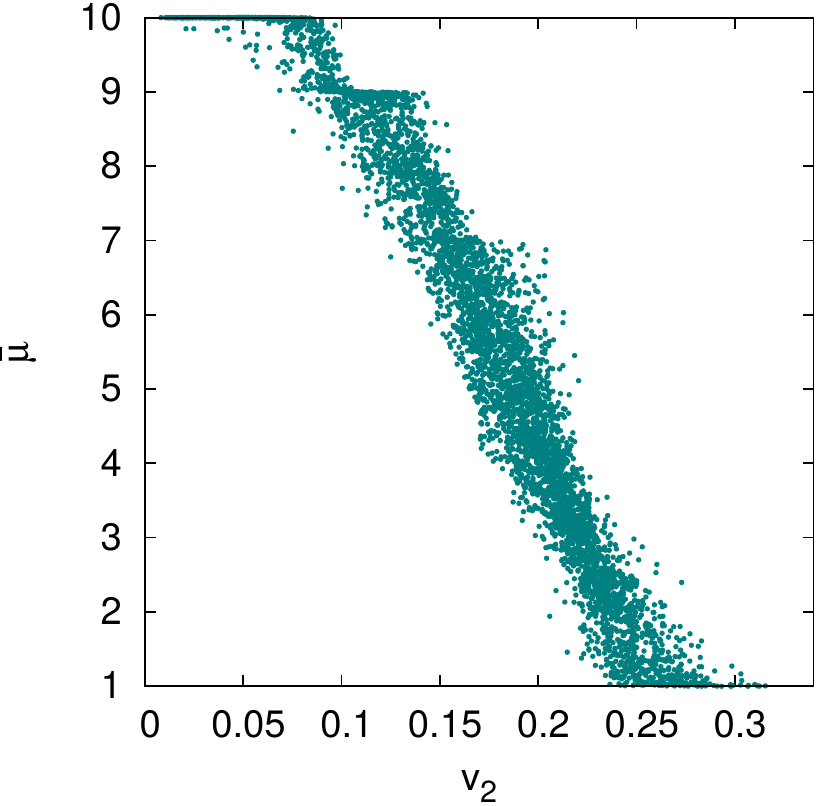}}
\caption{%
Top: correlation of the resulting sorting variable $\bar\mu$ with the initial ordering. Every 
dot represents one event.
Bottom: correlation of $\bar \mu$ with $v_2$ determined for each event via
the event plane method.
\label{f:muv2mu1}}
\end{figure}
In Fig.~\ref{f:muv2mu1} top one can see that 
the initial ordering indeed has nothing to do with the final ordering of events.
In the bottom panel of that Figure we show the correlation of final sorting variable
$\bar \mu$ with the value of $v_2$ determined in each event via the event-plane 
method (results from cummulant method are practically identical). One can clearly see that 
the ordering is given by the elliptic anisotropy of the particle distribution. 

Note that when we started the sorting algorithm with initial ordering according to 
the value of $q_2$ in each event, the final correlation was just a mirror image of 
Fig.~\ref{f:muv2mu1} bottom. Small values of $v_2$ corresponded to small $\bar \mu$
and high values of $v_2$ to high values of $\bar \mu$. This illustrates that the 
algorithm always converges to a sorting of events according to their similarity 
but the direction how they are ordered along $\bar \mu$ may be different. 

There is a congestion of events seen at $\bar \mu = 9$ and 10 in the upper panel of
Fig.~\ref{f:muv2mu1} and a step in the same place in the lower panel of that figure. 
This actually shows that the sorting was very clear for these two event bins. The congestion
is made out of events for which when evaluating $\bar \mu$ according 
to  eq.~(\ref{e:meanmu}), the sum has one clearly dominant term. The probabilities 
that the event might belong to other event bins are very small. 

\begin{figure*}
\centerline{\includegraphics[width=0.8\textwidth]{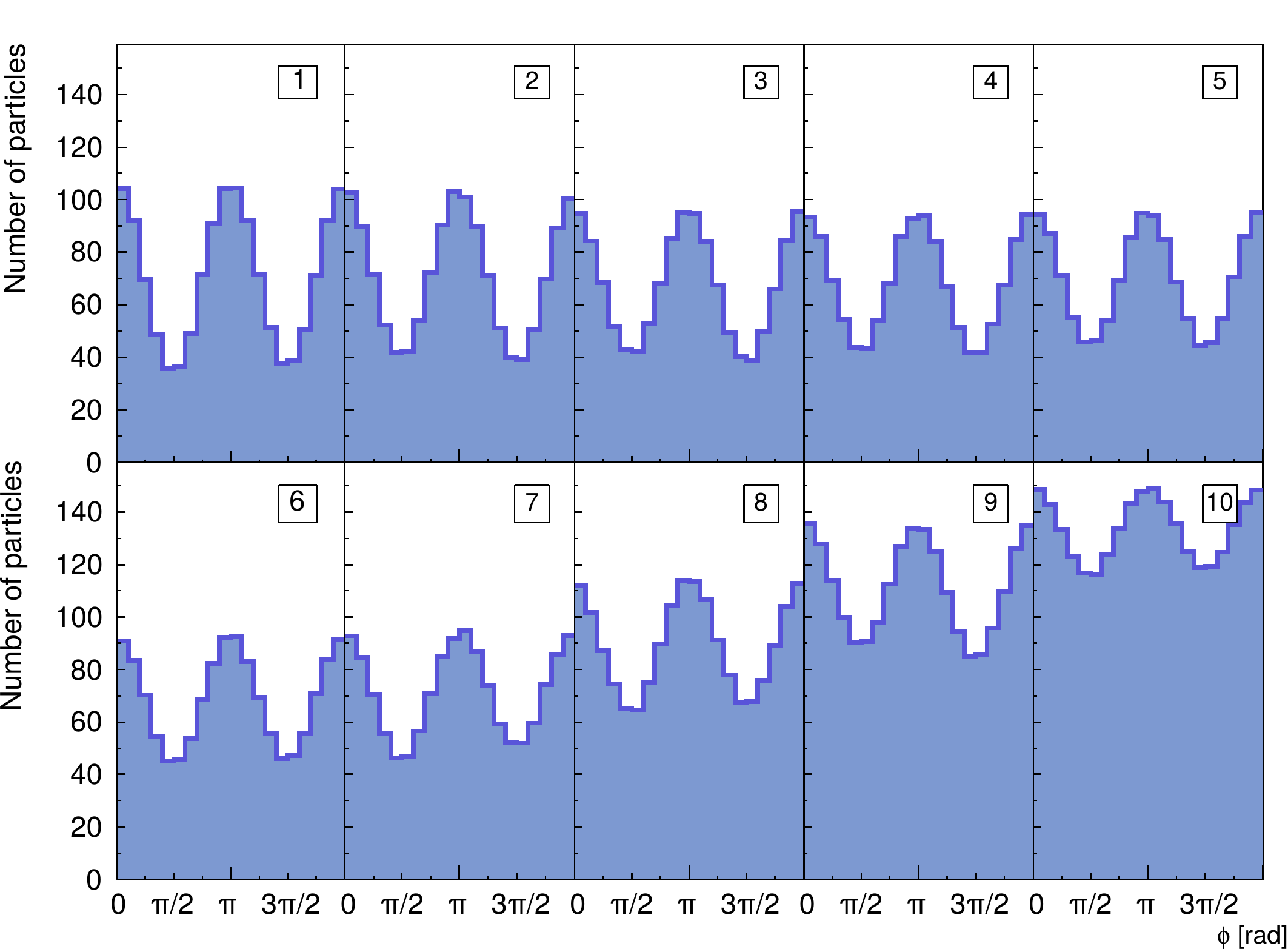}}
\caption{Average histograms of the 
azimuthal angles for event bins 1--10, with event bins indicated in the panels. 
Events with only $v_2$ and $v_1$. 
\label{f:histos1}}
\end{figure*}

For illustration, in Fig.~\ref{f:histos1} we show average histograms of the events in 
the individual event bins.  We see that the relative amplitude of the second order 
variation (\textit{i.e.} the amplitude divided by the mean value of the bins)
decreases from event bin 1 to event bin 10. 

The method is thus able to distinguish different shapes of hadron distributions 
and sort events according to them. In this simple case, this might not look like 
a big advantage as we could have sorted the events simply by measuring $v_2$. 
Therefore, we proceed with a more complicated example where the algorithm 
demonstrates its full power.


\subsection{Anisotropic flow}
\label{s:aniflow}

We parametrised the dependence of $v_n$'s on multiplicity through a fit to 
data from ALICE and ATLAS collaborations \cite{Gyulnara,ATLAS_vndata}. 
Coefficients $v_n$ for $n=1$ through 5 are parametrised according to eq.~(\ref{e:vnM}) 
with parameters summarised in Table~\ref{t:params}. 

Then we generated 5000 
events with multiplicities between 300 and 3000 pions and angular 
distributions according to 
\begin{equation}
P_5(\phi) = \frac{1}{2\pi} \left (  1 + 
\sum_{n=1}^5 2v_n \cos(n(\phi - \psi_n)) 
\right )\,   .
\end{equation}
The $v_n$'s for every event are set by   eq.~(\ref{e:vnM})  
and then smeared with Gaussian distribution with the width of 0.25. The phases 
$\psi_n$ are selected from uniform distribution and are not correlated with each other.

Now we have to address the question how to rotate the events so that the comparison
of individual events to the event bin histograms yields the most reasonable sorting. 
There are two symmetries at play here: rotational symmetry  and parity symmetry. We can
rotate an event around the collision axis, and we can also flip it so that
we get its mirror image. We have observed that both these 
symmetries influence the result. 

The two dominant components of flow anisotropy are second and third order. Hence, 
in our tests we focused on the corresponding event planes.  
If events are all aligned into the direction of second-order event plane, the algorithm 
becomes sensitive to the second-order anisotropy and to large extent ignores 
the third order. Analogically, alignment according to third-order event plane 
enhances the sensitivity to third order anisotropy. Thus the resulting sorting is rather 
sensitive to this choice. Furthermore, in both cases the algorithm 
distinguishes events which look like mirror images of each other 
(i.e.\ have opposite parity). 
This must be taken into account when designing the sorting algorithm.

%

The shape of hadron distribution is never solely determined by $v_2$ or by $v_3$.
It rather follows from their combination which is different in every event due to varying
phase difference of second and third order event plane. In order to take 
into account both these components of anisotropy we rotated all events according 
to the angle bisector between $\psi_2$ and $\psi_3$. We denote its azimuthal angle $\psi_{2-3}$.
Also, in order to take care of the parity symmetry the events were oriented so that 
$\psi_2$ is less than $\pi/2$ away from $\psi_{2-3}$ \emph{counterclockwise}.

The initial sorting of the events was random and the algorithm converged after 121
iterations. 

\begin{figure*}
\centerline{\includegraphics[width=0.42\textwidth]{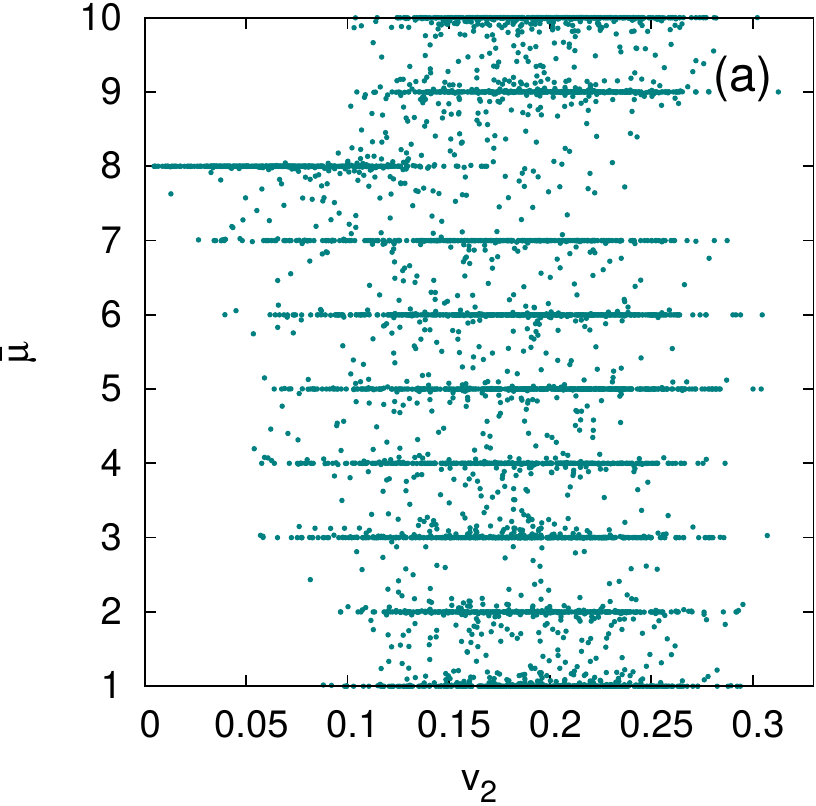}
\includegraphics[width=0.42\textwidth]{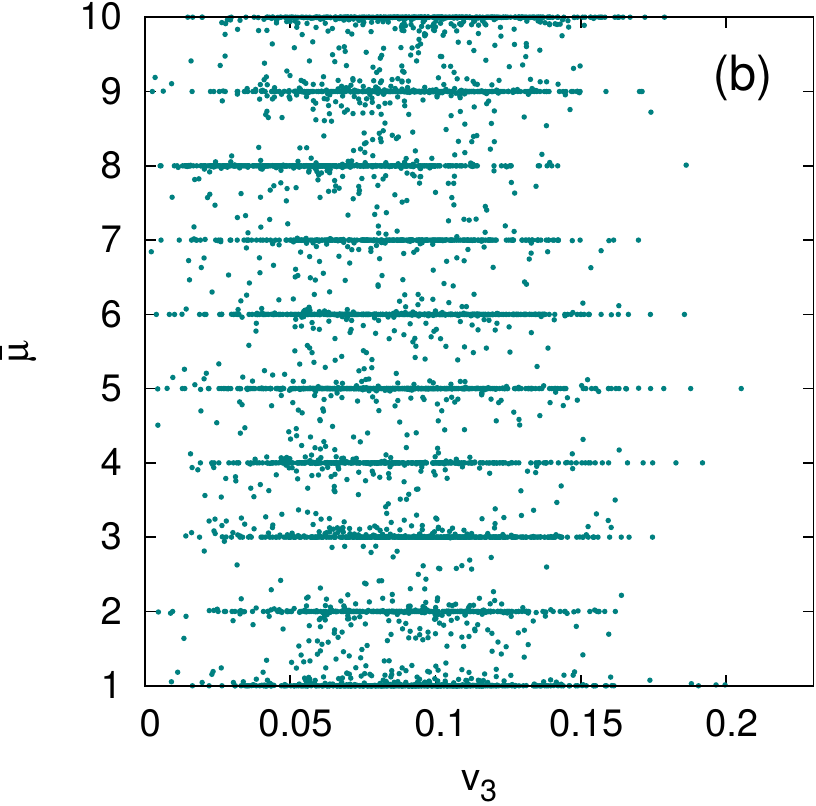}}
\centerline{\includegraphics[width=0.42\textwidth]{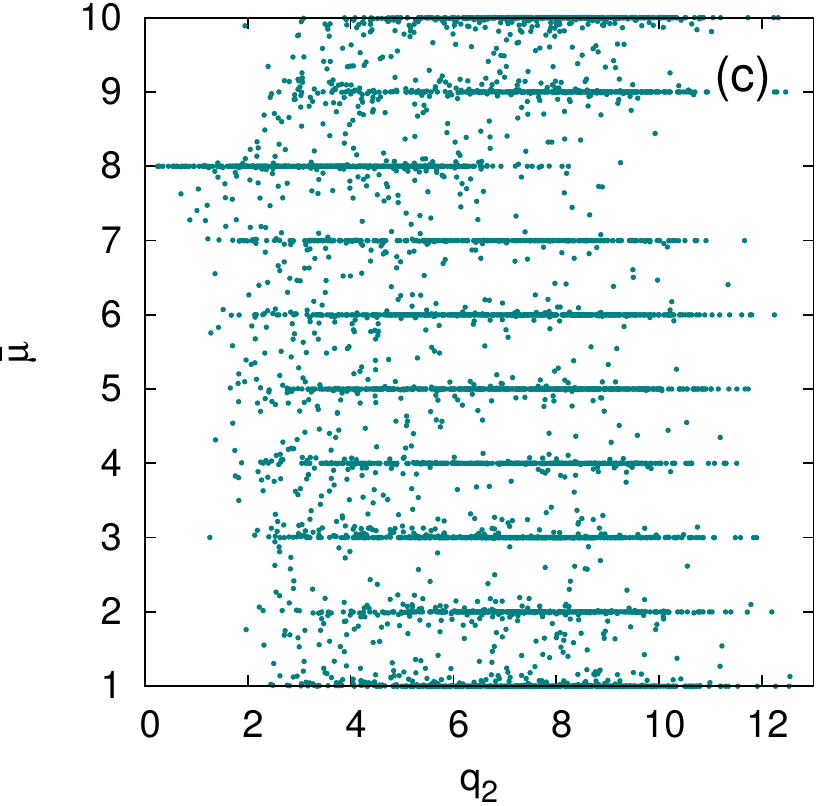}
\includegraphics[width=0.42\textwidth]{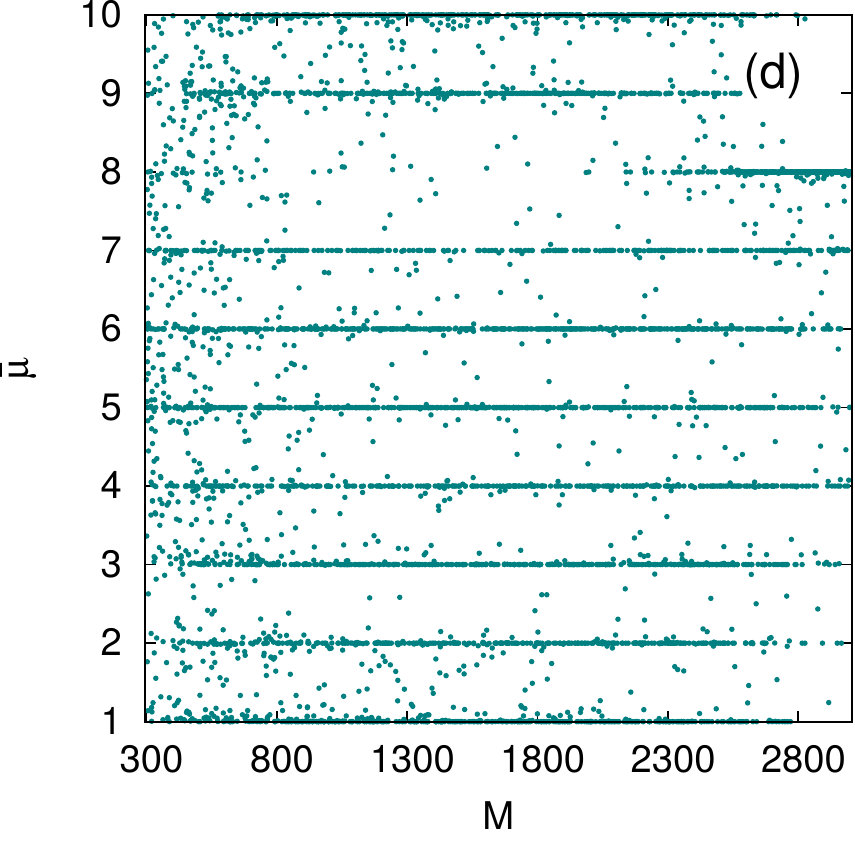}}
\caption{Correlation of various observables with final sorting variable $\bar\mu$. 
Simulated are events with anisotropies up to 5th order and initial rotation is 
according to $\psi_{2-3}$. Correlation with a) $v_2$, b) $v_3$, c) $q_2$, d) 
event multiplicity.
\label{f:corrvn}
}
\end{figure*}

From Fig.~\ref{f:corrvn} it is clearly seen that in this case sorting of the events is 
neither determined by $q_2$, nor by $v_2$, nor by $v_3$. 
Higher order terms also do not play a big role at all. 
The event shape is complex and results from 
an \emph{interplay} of all its simple characteristics. The message of the Figure is that 
$q_2$ may not be a good variable to select events according to their shape because 
panel (a) shows that it is not correlated with the overall shape of the event as soon as more 
flow harmonics are involved. 
Note that in our toy model there is neither correlation between the flow harmonics 
nor between the event planes of different order. This may not be so in real events  
and then the correlation between sorting variable $\bar \mu$ and some of the 
measured quantities may appear. What we show is thus rather an extreme case. It calls, 
however, for attention: the overall shape of an event and thus the evolution dynamics 
running in that event cannot be simplified into a single measured variable. There might 
be a counterargument that variables like e.g.\ $q_2$ which are used  in 
Event Shape Engineering are proved to be good at event sorting, because events 
with different values of $q_2$ show different values of other measured quantities. 
However,  in addition to this, our method 
optimizes the sorting so that events which are placed close 
to each other share as many event shape characteristics as possible. Figure~\ref{f:corrvn}
shows that such a sorting may not be connected with any of the commonly used variables. 

In Fig.~\ref{f:corrvn} we again observe that the resulting values 
of $\bar \mu$ are grouped around integers. This actually means that the 
assignment of the events into event bins is very clear, because in determining the value 
of $\bar \mu$ from eq.~(\ref{e:meanmu}) the probability $P(\mu|\{n_i\})$ 
is (close to) one for certain $\mu$ and (nearly) zero elsewhere. The event then 
clearly belongs to the event bin $\mu$.

When we tried to start the sorting algorithm with initial ordering according to the value of 
$q_2$ it failed to converge within reasonable time (5000 iterations). This happens sometimes
when unfavourable initial ordering is used. On the other hand, in other occasions we have
checked that also with different initial ordering of events the algorithm converged to
identical final sorting. The only difference could arise from the feature that the algorithm
does not follow any specific condition in which direction the events should be sorted. Thus different
initial orderings may end up in mutually reversed final orderings. 

\begin{figure*}
\centerline{\includegraphics[width=0.8\textwidth]{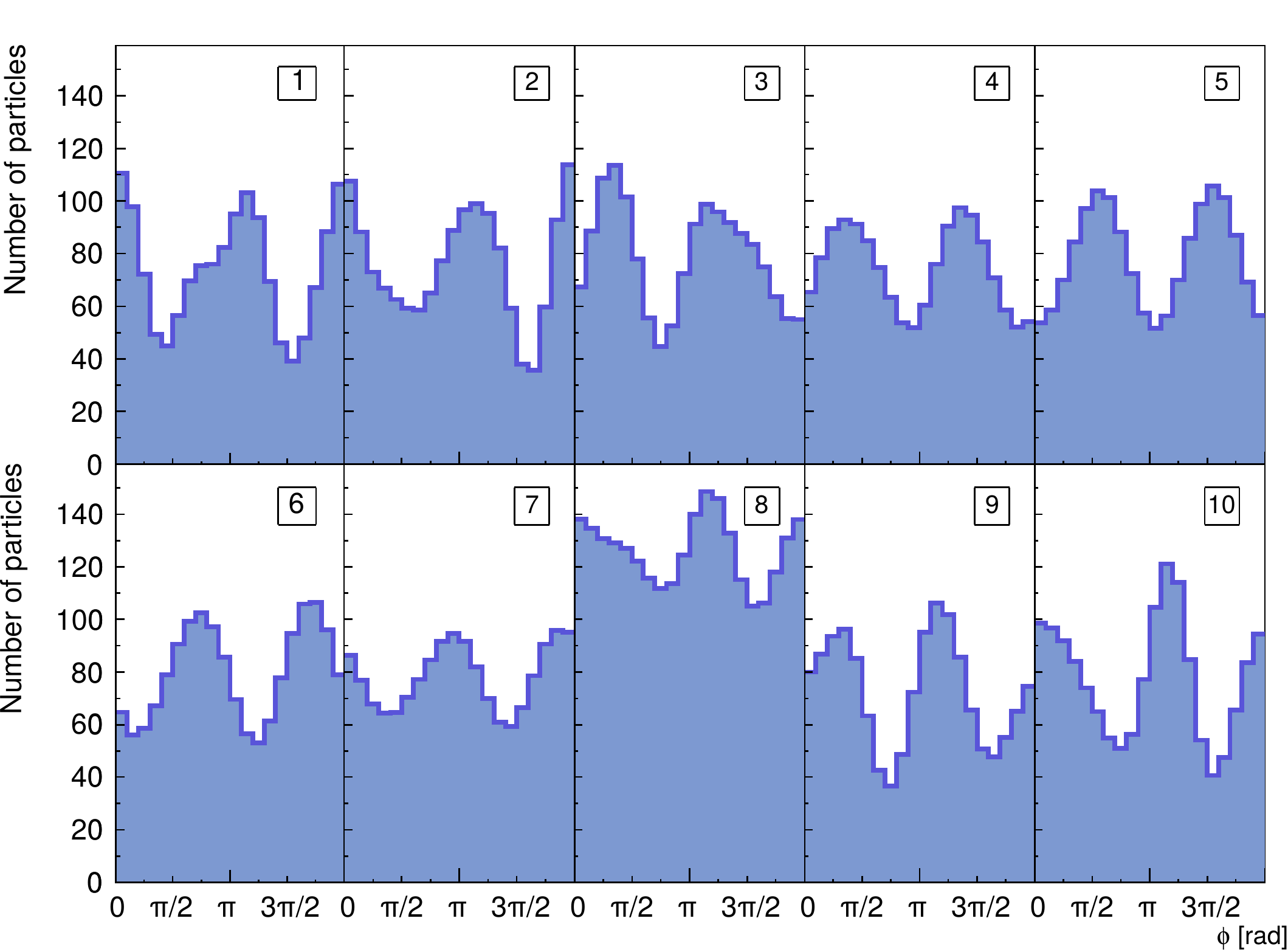}}
\caption{Average histograms of the 
azimuthal angles for event bins 1--10, with event bins indicated in the panels. 
Events with anisotropies up to 5th order.
\label{f:allnvhist}}
\end{figure*}

In Fig.~\ref{f:allnvhist} we show the resulting average angular histograms after the sorting. 
The events indeed differ by their \emph{shape}, not just only by the value of one of the 
flow harmonics. 
Since the second and third-order anisotropies are dominant and due to the initial 
rotation of the events, higher order harmonics are washed out by averaging over the event 
bins and not seen in the Figure. 

A question may appear to what extent the algorithm would sort the events even if they
would be drawn from the same distribution. It would still try to place closer together
events with similar histograms, and further apart those events with more different 
histograms.
Then---if in doubt---one can test the hypothesis that events are drawn from the same
distribution e.g.~with the method proposed in~\cite{Bityukov:2013dya}
or with the help of Kolmogorov-Smirnov test.


\section{Application to AMPT events}
\label{s:ampt}

After we have established and tuned the sorting algorithm, we now use it in a more 
realistic setting with events generated by the AMPT model \cite{AMPT}. 

The model was used as  commonly distributed with two modifications which
are recommended for  realistic simulation of Pb+Pb collisions at the LHC
collision energy $\sqrt{s_{NN}} = 2.76$~TeV: the parton screening mass was re-set to  
2.097~fm$^{-1}$  and the string melting was turned on in order to avoid the 
underestimation of partonic effects \cite{AMPT}.

We have generated 2000 events which correspond to the 0--20\% centrality class. On the 
generated particles we have applied rapidity cut $|y|<1$ in order to roughly simulate
the acceptance of central tracking detectors. For the first rough analysis we have 
taken all charged hadrons and run the sorting algorithm on them.

\begin{figure*}
\centerline{\includegraphics[width=0.42\textwidth]{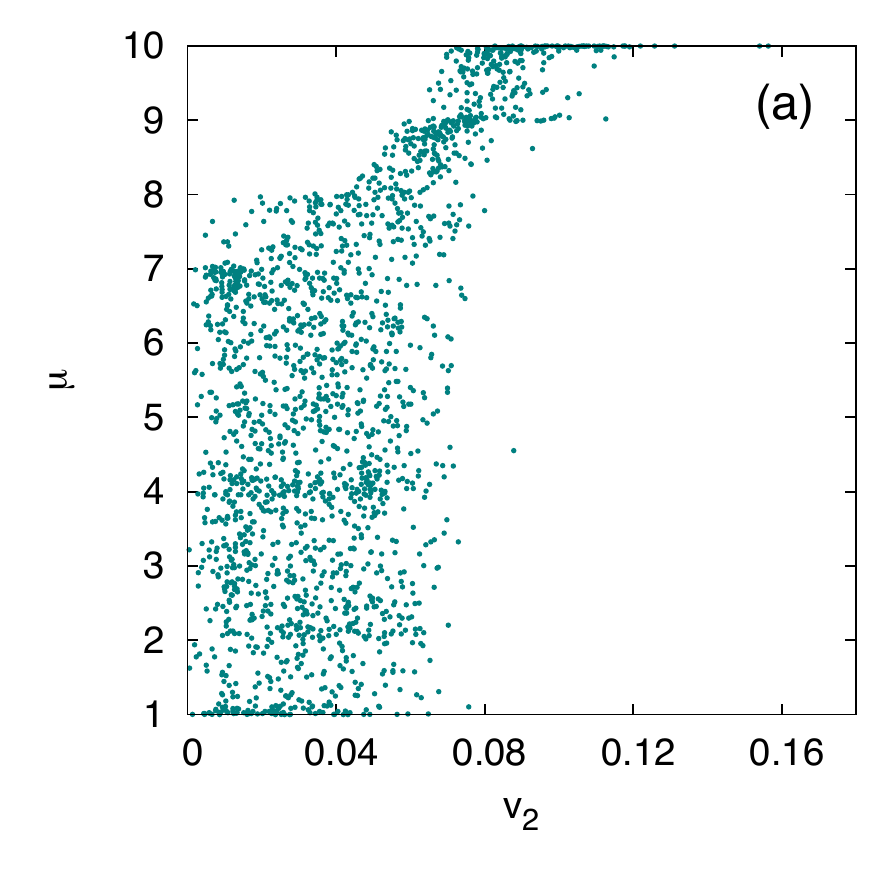}
\includegraphics[width=0.42\textwidth]{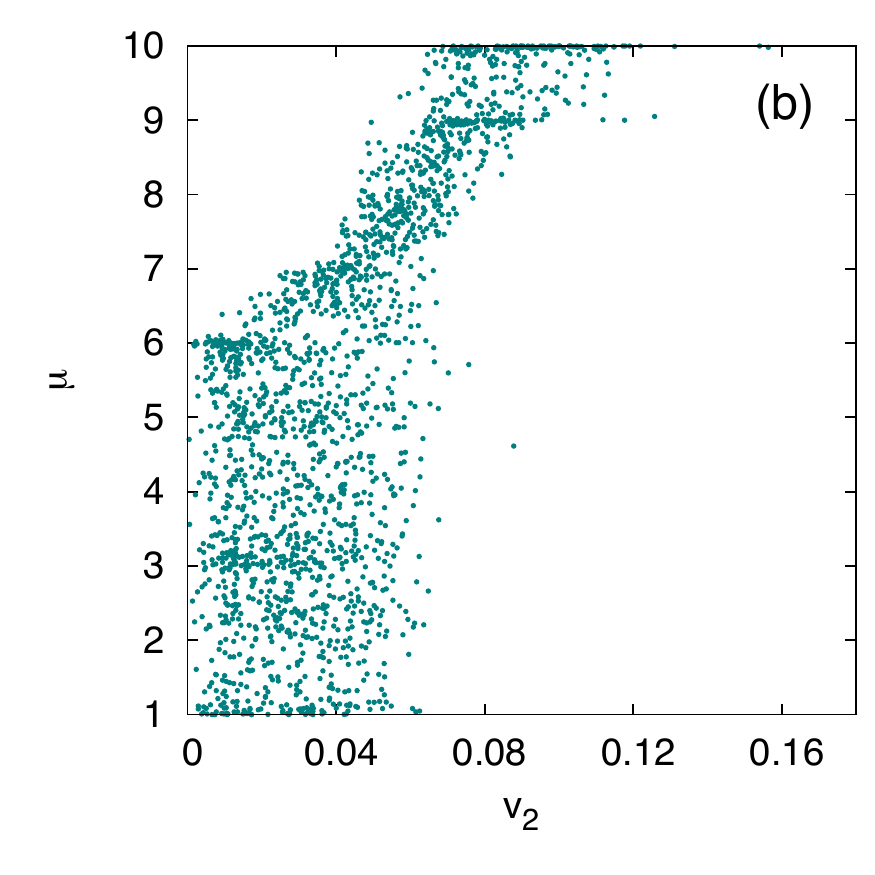}}
\centerline{\includegraphics[width=0.42\textwidth]{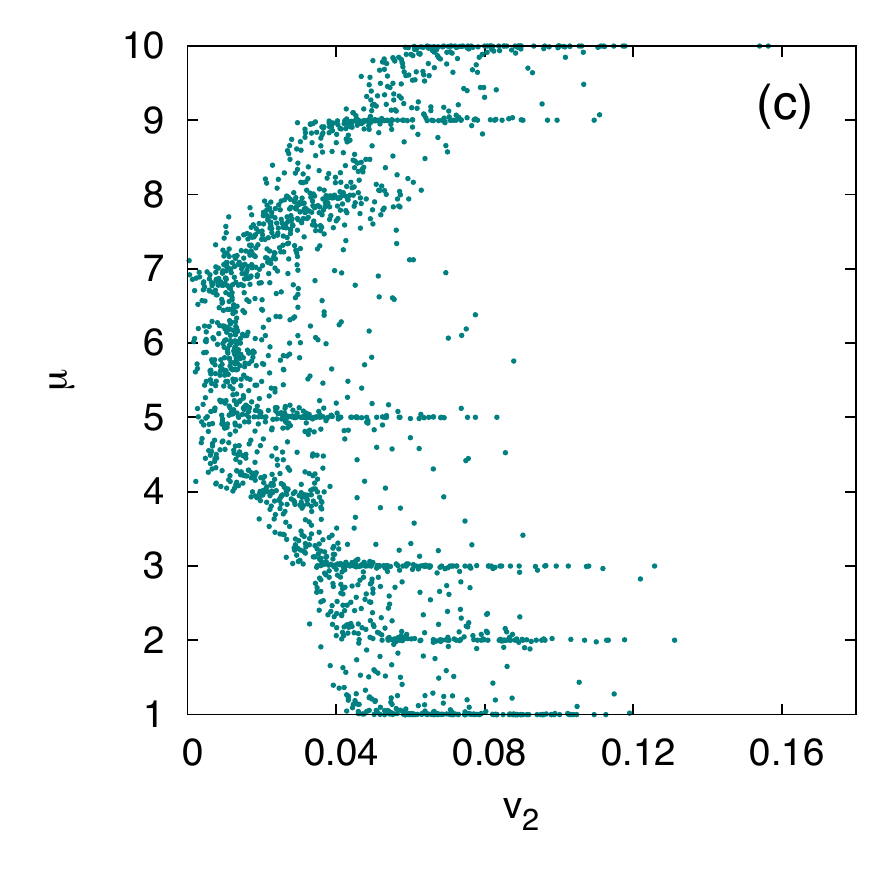}
\includegraphics[width=0.42\textwidth]{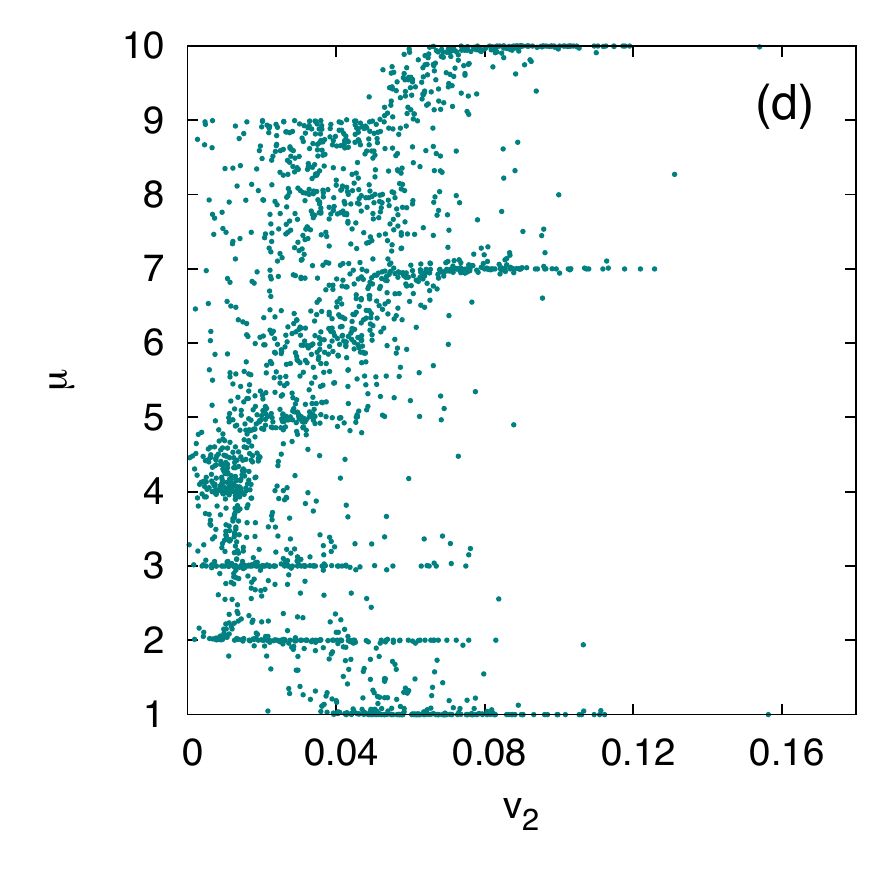}}
\caption{
\label{f:corrv2ampt}
Correlation of the final sorting variable $\bar \mu$ with $v_2$ of individual events for events 
generated by AMPT. a) Initial alignment of the events according to $\psi_2$, b) initial alignment 
according to $\psi_2$ and $v_2$ evaluated with $p_t$ weight, c) initial alignment according 
to $\psi_3$, d) initial alignment according to $\psi_{2-3}$. 
}
\end{figure*}

The event shapes are dominated by the second order anisotropy, but it is not the only feature 
that determines the shape of the fireball. When running our sorting algorithm, we have tried
three different initial alignments: according to  $\psi_2$, $\psi_3$ and $\psi_{2-3}$. 
In all three cases only $v_2$ and $q_2$ of the events show any pattern of correlation with the 
final sorting. This is shown in Fig.~\ref{f:corrv2ampt}.
%
%
Surprisingly, even in the case of $\psi_3$ initial alignment we have found no correlation 
of final ordering with $v_3$. On the other hand, as seen  in Fig.~\ref{f:corrv2ampt},
there is a pattern that shows that also here the sorting of events is strongly influenced 
by the second-order anisotropy. It will be interesting in the future to see if this dominance
of second-order anisotropy survives also in other centrality classes, particularly in more 
exclusively selected central event. 

It is also interesting to see that in realistic simulation, alignment with respect to $\psi_2$
does not  automatically lead to such a clear correlation of $v_2$ and $\bar \mu$ as it was 
the case with our toy model. This is seen in the upper two panels of Fig.~\ref{f:corrv2ampt}. 
Such a correlation exists for the two or three event bins with highest $\mu$'s. There,
higher $\mu$ corresponds to a higher value of $v_2$. However,  for 
$\bar\mu$ below 7 there seems to be no correlation between the ordering 
of the event and $v_2$. There is slightly more correlation in case weighted $v_2$ 
evaluation, as seen in Fig.~\ref{f:corrv2ampt}. We would like to understand where the 
difference between the events shapes comes from, which forces them into different event 
bins. 
\begin{figure*}
\centerline{\includegraphics[width=0.8\textwidth]{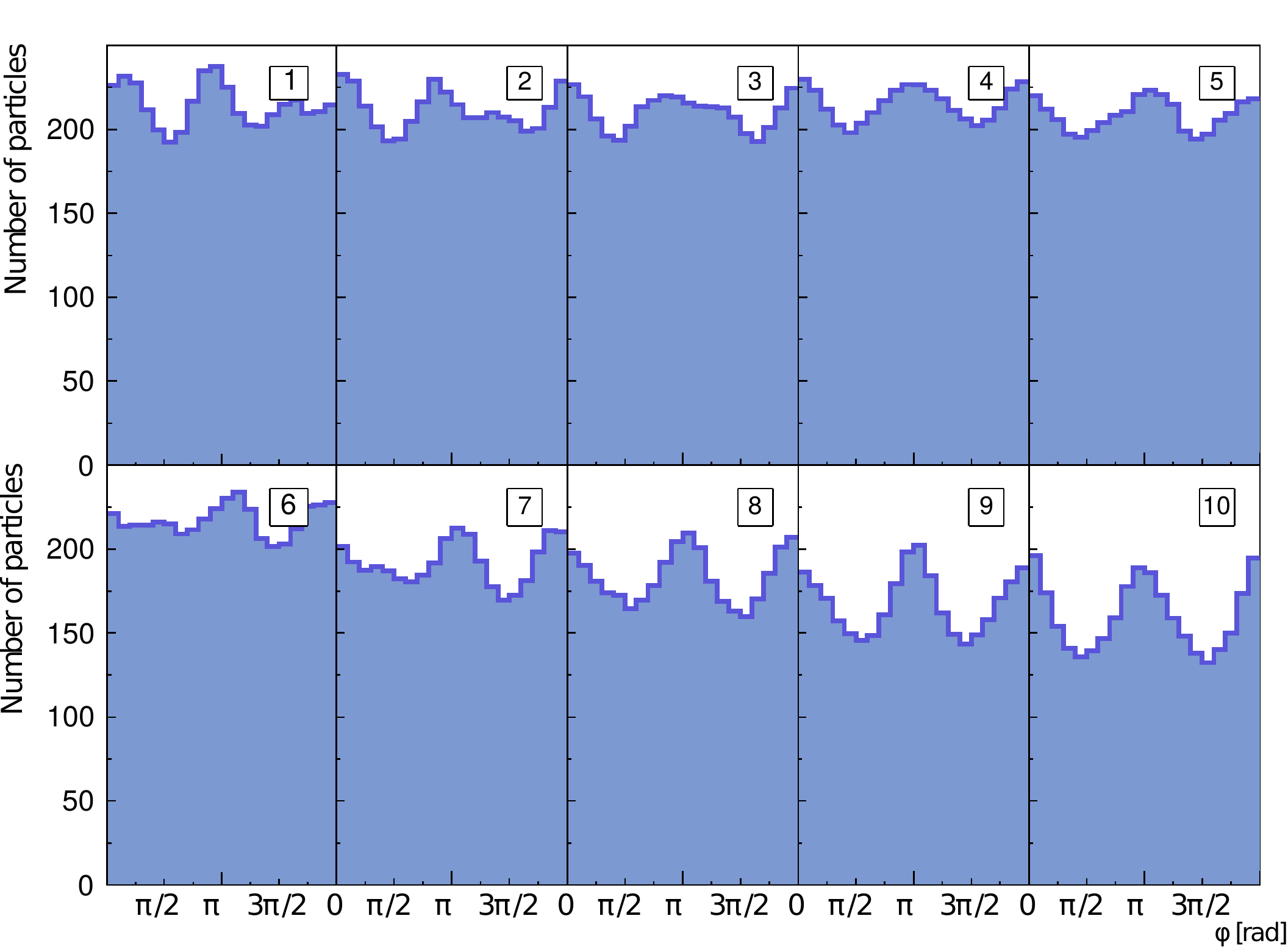}}
\caption{Average histograms of the 
azimuthal angles for event bins 1--10, from events generated with AMPT. Initial 
alignment of events according to $\psi_2$.
\label{f:amptafter}}
\end{figure*}
To this aim, we show in Fig.~\ref{f:amptafter} the average histograms in 
individual event bins after the sorting procedure has converged. In event bins 8, 9, and 10
the gradual growth of $v_2$ is evident. In the other event bins the histograms 
show a more complicated structure, where higher order anisotropies also give an important 
contribution. We recall, however, that no clear correlation of any higher $v_n$ with 
the obtained $\bar \mu$ is observed. 

\begin{figure}
\includegraphics[width=.42\textwidth]{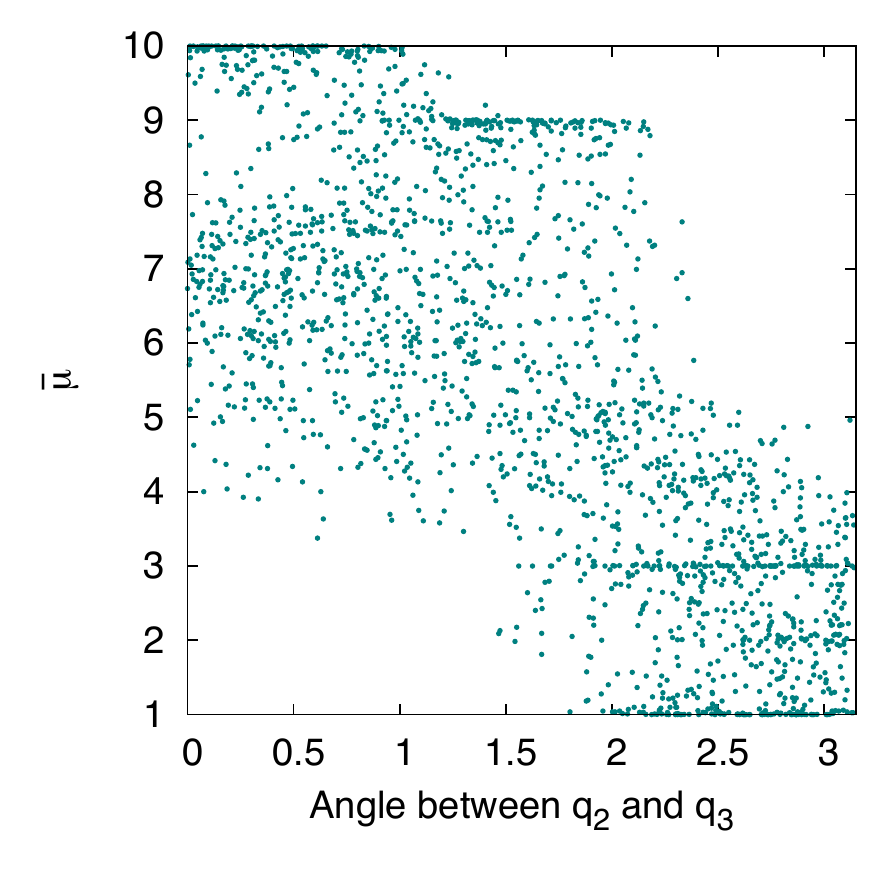}
\caption{The correlation of the  angle between the flow vectors
$\vec q_2$ and $\vec q_3$ and 
the average bin number of the event $\bar \mu$.  Events aligned according to 
$\psi_3$.
\label{f:amptq2q3}}
\end{figure}

The next suspected cause of the difference of  events is the relative angle between 
the directions of $\vec q_2$ and $\vec q_3$. We thus studied the correlation between 
$\bar \mu$ and the relative angle. In most cases, no correlation was observed. There 
is a hint of correlation, though, in case that the events are aligned according 
to $\psi_3$. This  is shown in Fig.~\ref{f:amptq2q3}.
Indeed, there seems to be a slight 
correspondence between the assignment to an event bin
and the angle between $\vec q_2$ and $\vec q_3$. 
The interpretation is at hand, that for the event shape the relative
phase is decisive. Unfortunately, this correlation is gone when we align the events 
differently. Thus, there are hints that the relative angle is important, but 
the assignment of 
an event to an event bin appears to be given by  more complicated interplay of 
various individual features.


\section{Conclusions and outlook}
\label{s:conc}

It is very useful to have a method able to sort events in such a way that it is possible 
to select those with very similar momentum distribution. One can then assume that 
they must have undergone similar evolution and this makes it possible to study 
the dynamics of hot expanding matter more exclusively. 

We tested this method on artificial events generated with AMPT. 
It showed that dividing the events into classes according one selected variable, 
usually $q_2$, does not really correspond to selecting event with the same shapes. 
It still seems that the main role in determining the event shape---at least for 
the studied centrality class 0--20\%---is played by the second-order anisotropy. 
Nevertheless, other features are important as well. In one case we could identify 
the difference $\psi_2-\psi_3$ to co-determine the assignment to event bins, but 
this observation is not universal for any initial event alignment and any way of 
evaluation of $q_n$'s.

In addition to the explanation of the method of Event Shape Sorting, 
we thus gave a first superficial study of realistic events with the proposed method.
The study of event shapes generated by AMPT for various centralities which would
include thorough analysis of all features that influence the final shape and their physics 
interpretation would go beyond the scope of this work and we plan to publish it 
in a separate paper. 

As was mentioned already in the Introduction, the presented method allows to select 
groups of events with similar momentum distributions. If used in data analysis, one can 
then measure various quantities on such events and study how they are related to the 
event shape.  

The method might even allow to get as close to single-event femtoscopy as 
possible. First, when doing femtoscopy with a single event one would run into 
difficulties with the uncorrelated reference distribution which is usually constructed 
through the event mixing technique. Event shape sorting could provide a selection of 
events with similar momentum distributions which would make a suitable sample for 
event mixing. Second, the statistics in a single event would be too small to perform a 3D 
analysis. However, one could take a sample of events with similar momentum distributions 
and reasonably expect that they also have the same sizes and undergo the same dynamics. 
Then, one could analyse the correlation function integrated over the whole selected 
event sample. The feasibility of such studies will be investigated in the future. 

An interesting application appears to be  the classification of events from U+U
collisions. Due to deformation of the colliding nuclei one expects large fluctuating
anisotropies of the transverse flow. Surprisingly, a preliminary study of azimuthally
sensitive correlation radii showed no dependence on the value of $q_2$
\cite{STAR_UUposter}. The latter
was employed for the selection of events with different final state anisotropy. A
short inspection of our Fig.~\ref{f:corrvn}c would suggest that this is no
surprise at all! As soon as there are other harmonic components of the anisotropy,
the shape of the events is more complex. The ``proper'' partition of events
into various classes by the type of anisotropy should be done differently. Our algorithm
can do such a proper classification.

Let us also comment again on the 
interesting though perhaps academic question, how the proposed algorithm 
would proceed if all events were indeed generated from the same underlying probability 
distribution and the only differences between them would be due to statistical 
fluctuations. The algorithm would be sensitive to the differences whatever their 
cause might be. Thus it would sort the events so that typical fluctuations within one 
event class would be below the normal statistical ones. Such a situation could be 
detected with the help of standard statistical tests, like e.g.\ the Kolmogorov-Smirnov test.  
In real data we do not expect this to happen, however. 

Also, there are still technical issues which require some discussion and 
will be addressed in the 
future. Most important is the ambiguity if initial rotation of the events for which we do not 
yet have optimised rules. Another is the rather high requirement on CPU time for even 
moderately large 
event samples. Note however, that we have not tried any fancy computational 
optimisation of the algorithm so far, 
and hence we would expect some room for improvement 
here. 

In fact, we also work on a well optimized routine that can readily be taken and applied 
directly in data analysis. Integration into standard packages like ROOT or
HistFitter  \cite{Baak:2014wma} will be addressed, as well.

In spite of this, we believe that the Event Shape Sorting is worthwhile to apply in real data 
analysis and carries potential to gain us better insight into nuclear dynamics in heavy ion 
collisions.

\section*{Acknowledgement}

We are thankful to Sergei Voloshin, J\"urgen Schukraft, Arkady Taranenko, 
and Burkhard K\"ampfer for valuable comments
and to Serguei Bityukov for pointing us to Ref.~\cite{Bityukov:2013dya}. 
BT thanks the Frankfurt Institute for Advanced Studies for warm hospitality
during his stay where a part of this study was completed.


\end{document}